\documentclass[reqno]{article}
\usepackage{alltt}
\usepackage{graphicx}
\title{Morphogenesis and dynamics of quantum state}
\author{Peter Leifer}
\date{Cathedra of Informatics, Crimea State Engineering and
Pedagogical University, \\
21 Sevastopolskaya st., 95015 Simferopol, Crimea, Ukraine; \\Hermon Laboratories, Ltd. 
Binyamina, 30500 Israel \\
leifer@bezeqint.net }
\begin{document}
\maketitle
\begin{abstract}
New construction of $4D$ dynamical space-time (DST) has been proposed in the framework of unification of relativity and quantum theory. Such unification is based solely on the fundamental notion of generalized coherent state (GCS) of N-level system and the geometry of unitary group $SU(N)$ acting in state space $C^N$. Neither contradictable notion of quantum particle, nor space-time coordinates (that cannot be a priori attached to nothing) are used in this construction. Morphogenesis of the ``field shell"-lump of GCS and its dynamics have been studied for $N=2$ in DST. The main technical problem is to find non-Abelian gauge field arising from conservation law of the local Hailtonian vector field. The last one may be expressed as parallel transport of local Hamiltonian in projective Hilbert space $CP(N-1)$. Co-movable local ``Lorentz frame" being attached to GCS is used for qubit encoding result of comparison of the parallel transported local Hamiltonian in infinitesimally close points. This leads to quasi-linear relativistic field equations with soliton-like solutions for ``field shell" in emerged DST. The terms ``comparison" and ``encoding" resemble human's procedure, but here they have objective content realized in invariant quantum dynamics. The dynamical motion of the lump in DST may be associated with ``kinesis" time whereas the evolution parameter describing morphogenesis of GCS evolving in $CP(N-1)$, may be naturally identified with ``metabole" time.
\end{abstract}

\section{Introduction}
The mystic role of time in everyday life, philosophy and even in modern physics is a good theme for amazing speculations. The nature of time is involved in theories of Big Bang, Multiverse, string extra-dimensions, etc. All these theories open vast area for scientific and scientific-like discussions. It is useful to clarify the root properties of time at least in some enough narrow area. I will discuss in this essay only physical aspect of this problem.

Even this restricted aspect is very wide. I try to show some modest, in comparison with mentioned above theories, approach to space-time problem in the framework of unification of relativity and quantum theory. This intrinsically geometric method of unification is yet not   approved theory but it may be treated as reasonable approach to future robust theory. It is interesting that the space-time arises here ``from inside" since quantum theory is built in background independent manner.

Quantum Universe in my model is represented by projective complex Hilbert space $CP(\infty)$ but for technical simplicity I anywhere use the $CP(N-1)$ assuming that GCS of N-level
system is taken as fundamental model of becoming quantum particles. The action states (superposition of quantum states with integer quanta of action) of this system has objective sense and their rays constitute $CP(N-1)$. All internal quantum dynamics develops in $CP(N-1)$ that serves as the base manifold of the tangent fibre bundle \cite{Le1,Le2}.

In order to get space-time structure one need the specific projection of the tangent fibre bundle. This projection is provided by attachment of co-moving ``Lorentz frame", originated by components of two infinitesimally close qubit spinors attached to smoothly moving GCS in $CP(N-1)$ under setup variation.

Attachment of the co-movable ``Lorentz frame" is similar to clock shows the phase of wave function in bright Feynman's simplified explanation of quantum electrodynamics \cite{Feynman1}.  Feynman discusses the \emph{amplitude of an event in stationary situation} since summation of amplitudes refers to fixed setup.
\begin{figure}[h]
  \includegraphics[width=1in]{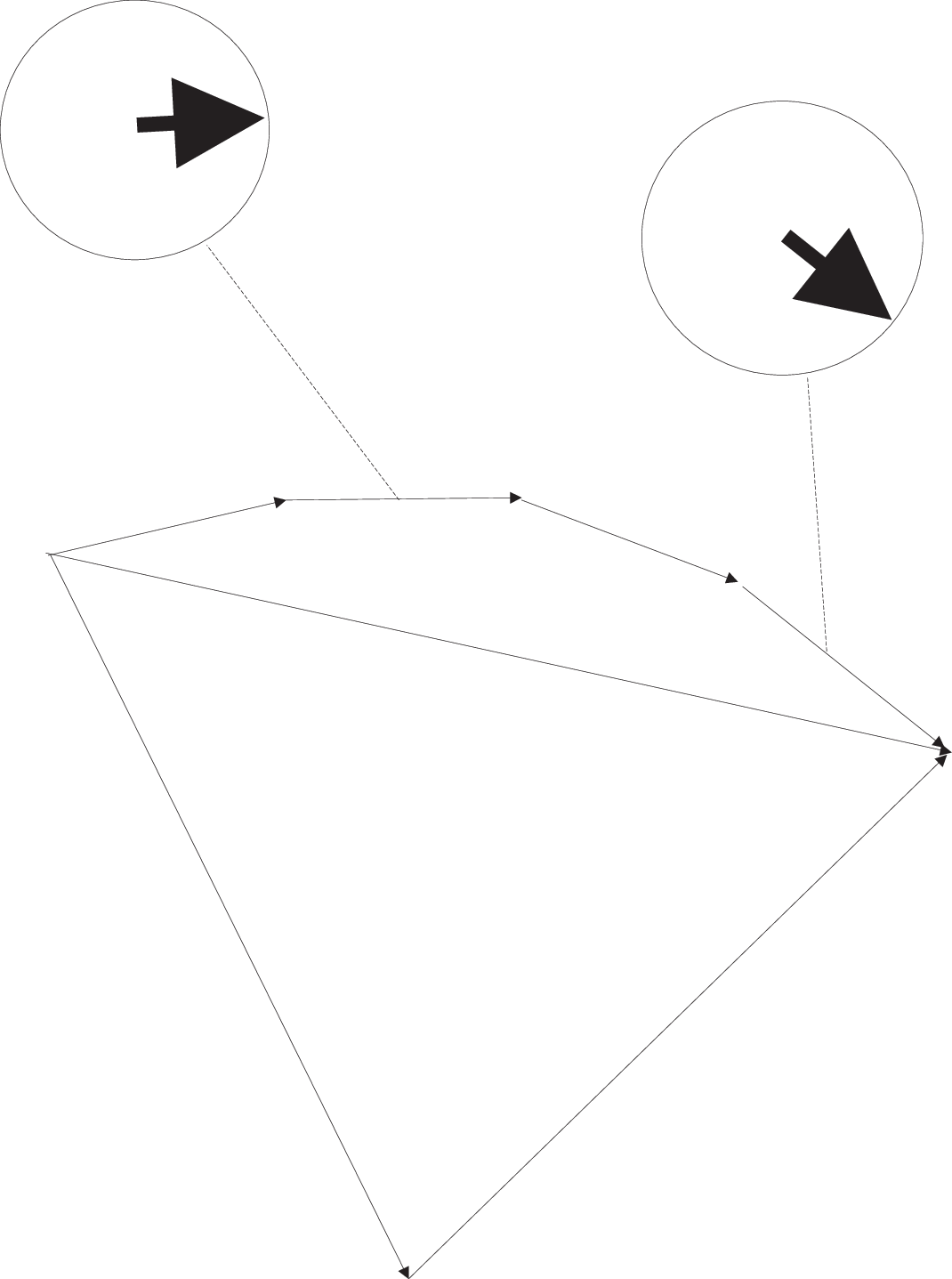}\\
  \caption{Fixed setup - summation and multiplication of state vectors of being particles. Feynman's summation of amplitudes corresponding to the time of light propagation from internal points of glass plate to detector. Equivalent amplitude arises as sum of ``forward" and ``backward" reflection from border surfaces.}\label{fig.1}
\end{figure}
Dynamical GCS moving due to setup variation requires operation with velocities of state deformation. This variable setup is described by ``field shell" that should dynamically conserve local Hamiltomian vector field \cite{Le2,Le3,Le4}.
I attached qubit spinor and further ``Lorentz frame" that define ``4-velocity" of some imaging point belonging to the DST.  This imaging point is some analog of Feynman's arrow but now in $4D$ DST. Quasi-linear partial differential equations arising as a consequence of conservation law of local Hamiltonian of evolving quantum system, define morphogenesis of non-Abelian (phase) gauge soliton-like ``field shell". So, we have a concentrated ``lump" associated with becoming quantum particle, see Figure 2.

\section{Why becoming quantum particles?}
Observable ``elementary" quantum particles are not classical material points. Success of Schr\"odinger is limited by non-relativistic quantum mechanics and it is in fact the last success of such kind. The notion of relativistic quantum particles is very contradictable and cannot serve as primordial construction in consistent quantum theory. I will use so-called N-level quantum system that is enough general in comparison with, say harmonic oscillator, and it is not so complicated as ``elementary particle". I will build
self-consistent construction of becoming ``field shell" of GCS together with DST.

\section{Separability and identification in relativity}
Space-time geometry was established during four revolutionary
steps: Euclidean axiomatization of 3-geometry, summarizing previous
\emph{mechanical hard body} measurements, Galileo-Newton's dynamics postulating
absolute space and time structure, Einstein's discovery of
pseudo-Euclidean space-time 4-geometry which physically based on
measurements by means of \emph{classical electromagnetic field}, and
Einstein's discovery of pseudo-Riemannean space-time 4-geometry
relies upon specification of same kind of measurements in
gravitation field. One sees that the modeling the space-time
geometry is boosted up by the development of measuring process in
more general physical conditions.

Quantum measurement has a long history but up to now it is badly defined
\cite{Penrose}. This is the reason why I propose a new procedure that is entirely
built in terms of geometry of $CP(N-1)$ \cite{Le1,Le2,Le3,Le4}. One of the most important property of this procedure is separability of physically different quantum states and a possibility to identify quantum system.

Generally, it is important to understand that the problem of identification of physically separated  objects is the root problem even in classical physics and that its recognition gave to Einstein the key to formalization of the relativistic kinematics and dynamics.
Indeed, only assuming the possibility to detect locally an approximate
coincidence of two pointwise bodies of a different nature it is
possible to build full kinematic scheme and the physical geometry of
space-time \cite{Einstein1,Einstein2}. As such the ``state" of the
local clock gives us local coordinates - the ``state" of the
incoming train. In the classical case the notions of the ``clock"
and the ``train" are intuitively clear and approximately may be identified with material points or even with space-time points (events). This supports the illusion that material bodies present in space-time (Einstein emphasized that it is not so!). Furthermore, Einstein especially notes that he did not discuss the inaccuracy of the
simultaneity of two {\it approximately coinciding events} that should
be overcame by some abstraction \cite{Einstein1}. This abstraction
is of course the neglect of finite sizes (and all internal degrees
of freedom) of the both real clock and train. It gives the
representation of these ``states" by mathematical points in
space-time. Thereby the local identification of two events is the
formal source of the classical relativistic theory. However quantum object requires especial embedding in space-time and its the identification with space-time point is impossible since the localization of quantum particles is state-dependent. Hence the identification of quantum objects requires a physically motivated operational procedure with corresponding mathematical description.

\section{Classification of quantum motions and local dynamical variables}
The local Cartan decomposition of the unitary group $SU(N)$ associated with initially chosen state vector  $|S>$ gives the invariant classification of unitary motions of the state vector and local dynamical variables (LDV's) represented by vector fields in complex projective Hilbert space $CP(N-1)$. Since any state $|S>$ has the isotropy group
$H=U(1)\times U(N)$, only the coset transformations $G/H=SU(N)/S[U(1)
\times U(N-1)]=CP(N-1)$ effectively act in $C^N$. Therefore the
ray representation of $SU(N)$ in $C^N$, in particular, the embedding
of $H$ and $G/H$ in $G$, is a state-dependent parametrization.
Technically the local $SU(N)$ unitary classification of the quantum motions requires the transition from the matrices of Pauli $\hat{\sigma}_{\alpha},(\alpha=1,...,3)$, Gell-Mann $\hat{\lambda}_{\alpha},(\alpha=1,...,8)$, and in general $N \times N$ matrices $\hat{\Lambda}_{\alpha}(N),(\alpha=1,...,N^2-1)$ of $AlgSU(N)$ to the tangent vector fields to $CP(N-1)$ in local coordinates \cite{Le1}.
Hence, there is a diffeomorphism between the space of the rays
marked by the local coordinates
\begin{equation}\label{6}
\pi^i_{(j)}=\cases{\frac{S^i}{S^j},&if $ 1 \leq i < j$ \cr
\frac{S^{i+1}}{S^j}&if $j \leq i < N-1$},
\end{equation}
in the map
 $U_j:\{|S>,|S^j| \neq 0 \}, j\geq 0$
and the group manifold of the coset transformations
$G/H=SU(N)/S[U(1) \times U(N-1)]=CP(N-1)$, and as well as the isotropy group of the corresponding ray.
This diffeomorphism is provided by the coefficient functions
\begin{equation}\label{17}
\Phi_{\sigma}^i = \lim_{\epsilon \to 0} \epsilon^{-1}
\biggl\{\frac{[\exp(i\epsilon \hat{\lambda}_{\sigma})]_m^i S^m}{[\exp(i
\epsilon \hat{\lambda}_{\sigma})]_m^j S^m }-\frac{S^i}{S^j} \biggr\}=
\lim_{\epsilon \to 0} \epsilon^{-1} \{ \pi^i(\epsilon
\hat{\lambda}_{\sigma}) -\pi^i \}
\end{equation}
of the local generators
\begin{equation}\label{18}
\overrightarrow{D}_{\sigma}=\Phi_{\sigma}^i \frac{\partial}{\partial \pi^i} + c.c.
\end{equation}
comprise of non-holonomic overloaded basis of $CP(N-1)$ \cite{Le1}.
This maps the unitary group $SU(N)$ onto the base manifold $CP(N-1)$ of the tangent fibre bundle. Now one may introduce Hamiltonian vector field as a tangent vector fields
\begin{equation}\label{19}
\overrightarrow{H}=\hbar \sum_{\alpha = 1}^{N^2 -1}\Omega^{\sigma}(\tau)D_{\sigma}=\hbar \sum_{\alpha = 1}^{N^2-1}\Omega^{\sigma}(\tau)\Phi_{\sigma}^i \frac{\partial}{\partial \pi^i} + c.c.
\end{equation}
whose coefficient functions $\Omega^{\sigma}(\tau)$ may be found under the condition of self-conservation of GCS expressed as affine parallel transport
\begin{equation}\label{20}
\frac{\delta (\Omega^{\sigma}(\tau)\Phi_{\sigma}^i) }{\delta \tau}= 0.
\end{equation}
of Hamiltonian vector field
$H^i=\Omega^{\sigma}(\tau)\Phi_{\sigma}^i$ agrees with Fubini-Study metric.

\section{Super-relativity}
The concept of super-relativity \cite{Le1,Le2} arose as development of
Fock's idea of ``relativity to measuring device" \cite{Fock}. This idea may be
treated originally as generalization of the relativity concept in space-time to
``functional relativity" in the state space \cite{Kryukov1,Kryukov2} under some reservations. However the power of Fock's program is limited in comparison with power of Einstein's concepts of special or general relativity. The main reason is that the notion of the ``measuring device" could not be correctly formulated in the own framework of the standard quantum theory. Some additional and, in fact, outlandish classical ingredients should be involved. It is well known ``measurement problem" in quantum theory \cite{Schroer}. In order to overcome this problem we should to clarify relations between state vector and dynamical variables of our N-level system.

It is very strange to think that state vector being treated as basic element of the \emph{full} description of quantum system does not influence on dynamical variable of quantum system. Formally quantum dynamical variable represented by hermitian operator in Hilbert space carrying representation of symmetry group, say, Lorentz group. All formal apparatus of quantum theory is based on the assumption that these operators depend only on the parameters of this group. Such approach reflects, say, the \emph{first order of relativity}: the physics is same if any \emph{complete setup} subject (kinematical, not dynamical!) shifts, rotations, boosts as whole in Minkowski space-time.
The question is: what happen if one slightly variates some device, say, rotates a filter or, better, changes magnetic field around dense flint \cite{Le4} in our complete setup? This variation leads to small variation of output state vector and may be associated with some state-dependent dynamical variable since output state depends upon incoming state and on the intensity of filter interaction with incoming state.

\section{Dynamical space-time as ``objective observer"}
I have assumed that the quantum measurement of the LDV being encoded with help infinitesimal Lorentz transformations of qubit spinor leads to emergence of the dynamical space-time that takes the place of the objective ``quantum measurement machine" formalizing the process of numerical encoding the results of comparisons of LDV's. Two these procedures are described below.

\subsection{LDV's comparison}
Local coordinates $\pi^i$ of the GCS in $CP(N-1)$ give reliable geometric tool for
the description of quantum dynamics during interaction or self-interaction. This
leads to evolution of GCS and that may be used in measuring process. Two essential components of any measurement are identification and comparison. The Cartan's idea of reference to the previous infinitesimally close GCS has been used. So one could avoid the necessity of the ``second body" used as a reference frame. Thereby, LDV is now a new essential element of quantum dynamics \cite{Le4}. We should be able to compare some LDV at two infinitesimally close GCS represented by points of $CP(N-1)$. Since LDV's are vector fields on $CP(N-1)$, the most natural mean of comparison of the LDV's is affine parallel transport
agrees with Fubini-Study metric \cite{Le1}.

\subsection{Encoding the results of comparison}
The results of the comparison of LDV's should be formalized by numerical encoding.
Thus one may say that ``LDV has been measured". The invariant encoding is based on
the geometry of $CP(N-1)$ and LDV dynamics, say, dynamics of the local Hamiltonian field.
Its affine parallel transport expresses the self-conservation of quantum lump associated with ``particle". In order to build the qubit spinor $\eta$ of the quantum question
$\hat{Q}$ \cite{Le5} two orthogonal vectors $\{|N>,|\Psi>\}$ have been used. Here $|N>$ is the complex normal and $|\Psi>$ tangent vector to $CP(N-1)$.
I will use following equations
\begin{eqnarray}\label{24}
\eta=\left(
  \begin{array}{cc}
    \alpha_{(\pi^1,...,\pi^{N-1})}  \\
    \beta_{(\pi^1,...,\pi^{N-1})} \\
  \end{array}
\right) = \left(
  \begin{array}{cc}
    \frac{<N|\hat{H}|\Psi>}{<N|N>}  \\
    \frac{<\Psi|\hat{H}|\Psi>}{<\Psi|\Psi>} \\
  \end{array}
\right)
\end{eqnarray}
for the
measurement of the Hamiltonian $\hat{H}$ at corresponding GCS.
Then from the infinitesimally close GCS
$(\pi^1+\delta^1,...,\pi^{N-1}+\delta^{N-1})$, whose shift is
induced by the interaction used for a measurement, one get a close
spinor $\eta+\delta \eta$ with the components
\begin{eqnarray}\label{25}
\eta + \delta \eta =\left(
  \begin{array}{cc}
    \alpha_{(\pi^1+\delta^1,...,\pi^{N-1}+\delta^{N-1})}  \\
    \beta_{(\pi^1+\delta^1,...,\pi^{N-1}+\delta^{N-1})} \\
  \end{array}
\right) = \left(
  \begin{array}{cc}
    \frac{<N|\hat{H'}|\Psi>}{<N|N>}  \\
    \frac{<\Psi|\hat{H'}|\Psi>}{<\Psi|\Psi>}
  \end{array}
\right).
\end{eqnarray}
Here $\hat{H}=\hbar \Omega^{\alpha}\hat{\lambda}_{\alpha}$ is the lift of Hamiltonian tangent vector field $H^i=\hbar \Omega^{\alpha} \Phi^i_{\alpha}$ from $(\pi^1,...,\pi^{N-1})$ and $\hat{H'}=\hbar(\Omega^{\alpha}+\delta \Omega^{\alpha}) \hat{\lambda}_{\alpha}$ is the lift of the same tangent vector field parallel transported from the infinitesimally close point
$(\pi^1+\delta^1,...,\pi^{N-1}+\delta^{N-1})$ back to the
$(\pi^1,...,\pi^{N-1})$ into the adjoint representation space.

Each quantum measurement consists of the procedure of encoding
of quantum dynamical variable into state of a ``pointer" of ``macroscopic measurement machine" \cite{Wezel}. Quantum lump takes the place of such extended ``pointer".
This extended pointer may be mapped onto dynamical space-time if one assumes
that transition from one GCS to another is accompanied by dynamical
transition from one Lorentz frame to another attached to adjacent point of the ``pointer", see Figure 2.
Thereby, infinitesimal Lorentz transformations define small
``dynamical space-time'' coordinates variations. It is convenient to take
Lorentz transformations in the following form
\begin{eqnarray}\label{28}
ct'=ct+(\vec{x} \vec{a}) \delta \tau \cr
\vec{x'}=\vec{x}+ct\vec{a} \delta \tau
+(\vec{\omega} \times \vec{x}) \delta \tau
\end{eqnarray}
where I put
$\vec{a}=(a_1/c,a_2/c,a_3/c), \quad
\vec{\omega}=(\omega_1,\omega_2,\omega_3)$ \cite{G} in order to have
for $\tau$ the physical dimension of time. The expression for the
``4-velocity" $ V^{\mu}$ is as follows
\begin{equation}\label{29}
V^{\mu}=\frac{\delta x^{\mu}}{\delta \tau} = (\vec{x} \vec{a},
ct\vec{a}  +\vec{\omega} \times \vec{x}) .
\end{equation}
The coordinates $x^\mu$ of imaging point in dynamical space-time serve here merely for
the parametrization of the energy distribution in the ``field
shell'' arising under ``morphogenesis" described by quasi-linear field
equations \cite{Le2,Le6,Le7}.

Any two infinitesimally close spinors $\eta$ and $\eta+\delta
\eta$ may be formally connected with infinitesimal ``Lorentz spin transformations
matrix'' \cite{G}
\begin{eqnarray}\label{31}
L=\left( \begin {array}{cc} 1-\frac{i}{2}\delta \tau ( \omega_3+ia_3 )
&-\frac{i}{2}\delta \tau ( \omega_1+ia_1 -i ( \omega_2+ia_2)) \cr
-\frac{i}{2}\delta \tau
 ( \omega_1+ia_1+i ( \omega_2+ia_2))
 &1-\frac{i}{2}\delta \tau( -\omega_3-ia_3)
\end {array} \right).
\end{eqnarray}
I have assumed that there is not only formal but dynamical reason for such transition
when Lorentz reference frame moves together with GCS.
Then ``quantum accelerations" $a_1,a_2,a_3$ and ``quantum angle velocities" $\omega_1,
\omega_2, \omega_3$ may be found in the linear approximation from
the equation
\begin{eqnarray}\label{32}
\eta+\delta \eta = L \eta
\end{eqnarray}
as functions of the qubit spinor components of the quantum question
depending on local coordinates $(\pi^1,...,\pi^{N-1})$.
\begin{figure}[h]
  \includegraphics[width=1in]{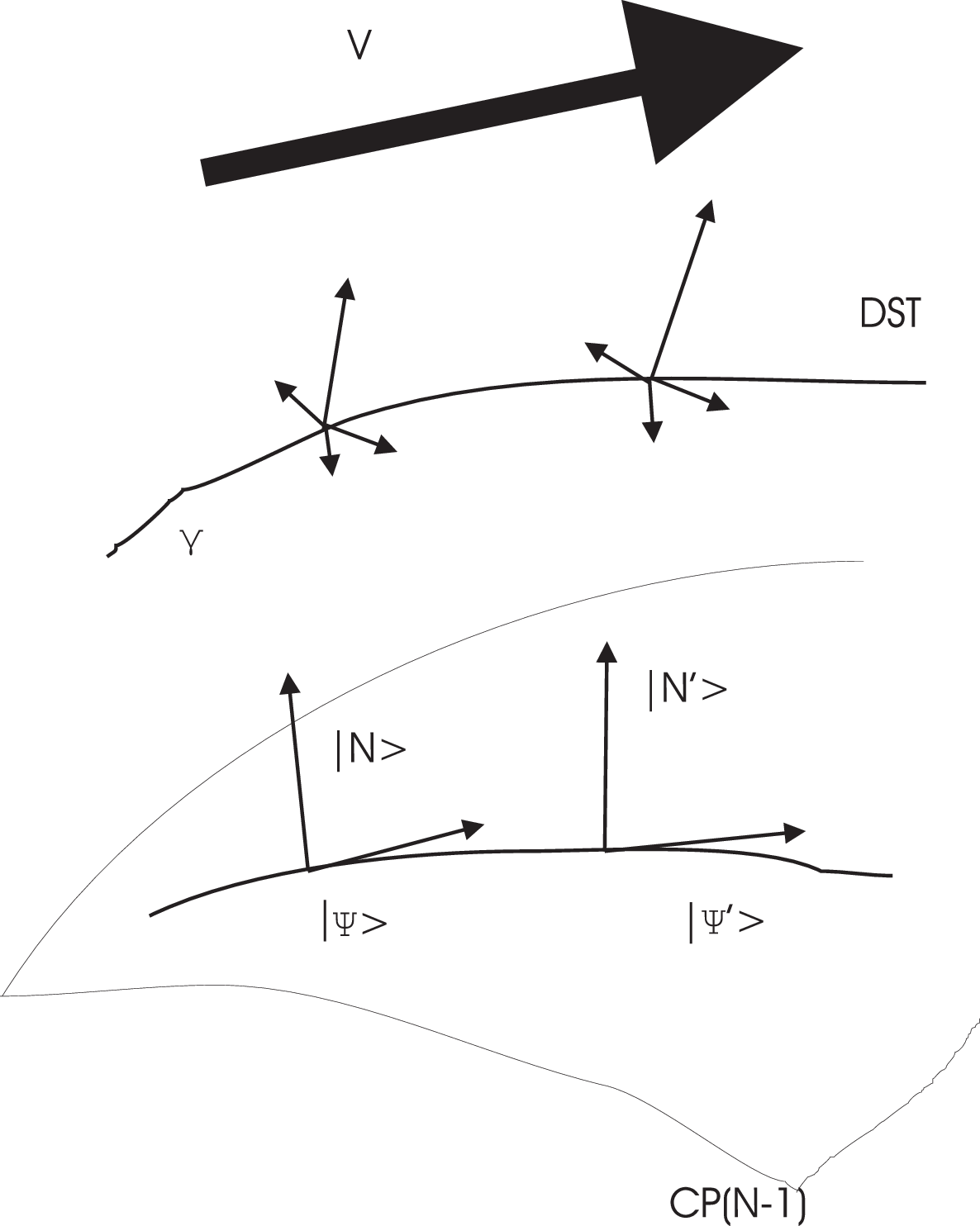}\\
  \caption{Dynamical setup for becoming lump - operations with LDV. In order to get effective sum of non-Abelian phases of $SU(N)$ transformation shaping the lump, one should integrate quasi-linear partial differential equations. The ``4-velocity" $V$ of imaging point in DST is parameterized by boosts and angle velocities of co-moving ``Lorentz reference frame" attached to trajectory in $CP(N-1)$.}\label{fig.2}
\end{figure}
\section{Morphogenesis of the lump}
Globally broken Lorentz symmetry widely discussed now (see, say, \cite{Kostel}), should be locally restored with help the affine parallel transport  of the local Hamiltonian in the projective Hilbert state space that leads to extended soliton-like solutions \cite{Le2}. It is defined by the velocity of variation of qubit spinor $\eta$ during parallel transport of local Hamiltonian. Moreover, there is some affine gauge field which in some sense restores global Lorentz invariance since the filed equations for the lump are relativistically invariant.  In fact not any classical field in space-time correspond to the parallel transport in $CP(N-1)$, but in dynamical space-time permissible only fields corresponding to conservation laws in $CP(N-1)$. These conservation laws are expressed by the affine parallel transport. The parallel transport of the local Hamiltonian provides the ``self-conservation" of extended object, i.e. the affine gauge fields couple the soliton-like system (\ref{40}) discussed in \cite{Le2,Le3}.

The field equations for the $SU(N)$ parameters $\Omega^{\alpha}$
dictated by the affine parallel transport of the Hamiltonian vector field
$H^i=\hbar \Omega^{\alpha}\Phi^i_{\alpha}$ (5)
read as quasi-linear PDE
\begin{equation}\label{40}
\frac{\delta \Omega^{\alpha}}{\delta \tau} = V^{\mu} \frac{\partial \Omega^{\alpha}}{\partial x^{\mu} } = -
(\Gamma^m_{mn} \Phi_{\beta}^n+\frac{\partial
\Phi_{\beta}^n}{\partial \pi^n}) \Omega^{\alpha}\Omega^{\beta},
\quad \frac{d\pi^k}{d\tau}= \Phi_{\beta}^k \Omega^{\beta}.
\end{equation}
  The PDE equation obtained as a consequence of the parallel transport of the local Hamiltonian for two-level system living in $CP(1)$ has been shortly discussed \cite{Le2,Le6,Le7}
\begin{equation}\label{41}
\frac{r}{c}\psi_t+ct\psi_r=F(u,v) \rho \cos \psi.
\end{equation}
The one of the exact solutions of this quasi-linear PDE is
\begin{eqnarray}\label{42}
\psi_{exact}(t,r)= \arctan \frac{\exp(2c\rho F(u,v)
f(r^2-c^2t^2))(ct+r)^{2F(u,v)}-1}{\exp(2c\rho F(u,v)
f(r^2-c^2t^2))(ct+r)^{2F(u,v)}+1},
\end{eqnarray}
where $f(r^2-c^2t^2)$ is an arbitrary function of the interval. These field equations describes energy distribution in the lump which does not exist a priori but is becoming during the self-interaction, see Figure 3.
\begin{figure}[h]
  \includegraphics[width=2in]{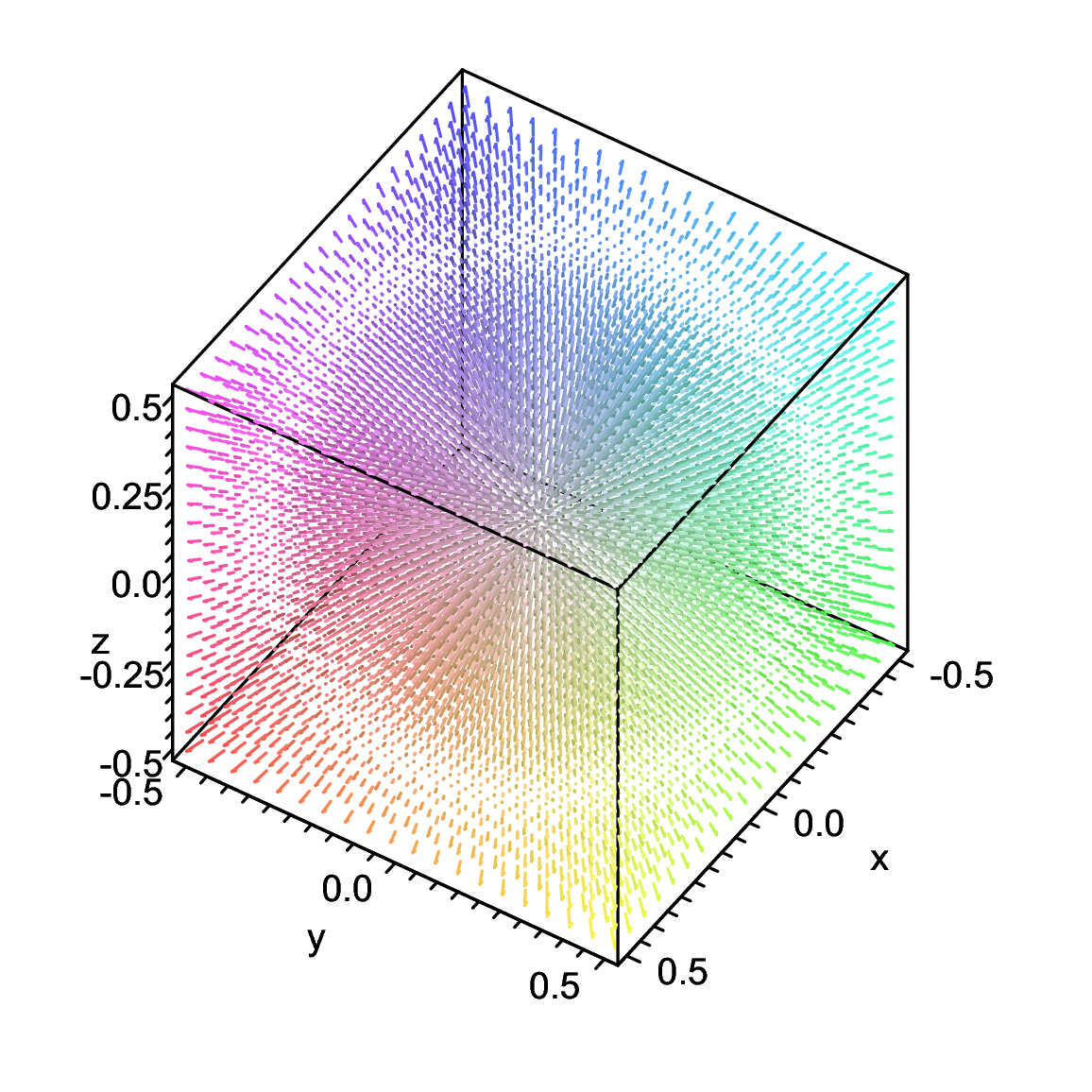}\\
  \caption{The non-monotonic distribution of the force field
in the lump looks like a bubble in the dynamical space-time}\label{fig.3}
\end{figure}
The question about stability of this solution and whether such approach deletes the necessity of some additional stabilization forces should be studied carefully.

\section{Conclusion}
Quantum theory of field (extended) objects without a priori
space-time geometry has been represented. Phase space
$CP(N-1)$ are used instead of space-time. The fate of
quantum system modeled by the generalized coherent states is rooted
in this manifold. Dynamical (state-dependent) space-time arises only
at the stage of the quantum ``yes/no" measurement. The quantum
measurement of the gauge ``field shell'' of the generalized coherent
state is described in terms of the affine parallel transport of the
local dynamical variables in $CP(N-1)$.

Now I see that my efforts to clarify space-time structure may evoke sardonic smile
since there are attempts ``forget time" at all \cite{Carlo}. I do not have now SPACE and TIME to analyze such approach in details, but I would like to note following.

1. If one accepts the concept of the Quantum Universe as space containing objective quantum states of matter, then not only time but the whole space-time is merely specific projection based on qubit encoding measurement results. One may say that projection does not have objective sense and therefore one may ``forget space-time". But there are objective generally \emph{flexible} relationships between physical values that should be discovered my measurement. Measurement is the procedure of a comparison of dynamical variables and encoding process of the comparison intended to give a number. Since physicist uses formal logic and ordinary mathematics based on ``yes/no" questions for comparison and encoding, there is a possibility (probably, not necessity!) to express all procedure in space-time terms.

2. My construction of GCS morphogenesis and dynamics is one of the possible realizations of the ``space-timeless language". Unfortunately complicated procedure of comparison and encoding is used in order to get space-time shape of localized lump. I do not like the part of construction that uses embedding projective space into flat Hilbert space. I prefer internal geometry formulation, but I do not see how we may reach  it; this topic requires additional investigation. Besides this the complicated structure of the self-consistent system of quasi-linear partial differential equations describing lump for $N>2$ is real challenge for future investigations.

3. It is interesting that dynamical space-time has a ``granular structure" \cite{Kostel} and agreement between different ``granule" requires physically motivated procedure. One may treat the DST as ``objective observer" with hope that our cognitive functions are more or less good adapted to it.

4. Evolution parameter $\tau$ in $CP(N-1)$ takes the place of Aristotle's ``metabole" time since it describes the temp of setup variation and it parameterizes the deformation  (morphogenesis) of generalized coherent state in $CP(N-1)$. On the other hand the motion of ``field shell" in the dynamical space-time runs according to relativistic ``kinesis" time.

\end{document}